\documentclass[12pt]{article}

\newcommand{\beq}{\begin{equation}}
\newcommand{\eeq}{\end{equation}}
\newcommand{\beqa}{\begin{eqnarray}}
\newcommand{\eeqa}{\end{eqnarray}}
\newcommand{\beqar}{\begin{eqnarray*}}
\newcommand{\eeqar}{\end{eqnarray*}}

\def \la {\langle}
\def \ra {\rangle}

\begin{document}
\input epsf
\title{\bf  \large Measurements of semi-local and non-maximally
entangled states}

\author{
Berry Groisman and  Benni Reznik
{\ }\\
{\em \small School of Physics and Astronomy, Tel Aviv
University, Tel Aviv 69978, Israel}
}

\date{ }
\maketitle


\begin{abstract}
Consistency with relativistic causality narrows down dramatically
the class of measurable observables.
We argue that by weakening the preparation role of 
ideal measurements, many of these observables become measurable.
Particularly, we show 
by applying entanglement assisted remote operations, 
that all Hermitian observable of a $2\times2$-dimensional 
bipartite system are measurable. 
\end{abstract}

\section{Introduction}

Most of the states in the Hilbert space ${\cal H}_{AB}={\cal H}_A\otimes
{\cal H}_B$ of two systems (a bipartite system)
are entangled states.  Similarly, the eigenstates of most observables
that act on ${\cal H}_{AB}$, are entangled states.
This poses no problem as long as the systems $A$ and $B$ are both localized
in the same region of space -- any observable can in principle be
measured by some joint interaction between the systems A and B and
a localized measuring device.
However, once we move A and B apart to separate locations in space,
we must introduce a new  constraint: the measurements
should be consistent with relativistic causality. 
Hence, only the causal observables, which do not give rise 
to superluminal effects,  are physical.
This restriction to causal observables,
turns out to narrows down dramatically the class of
measurable observables.
For a pair of two-level systems (qubits) this seems to leave us
only the observables whose eigenstates are the trivial non-entangled
basis $(|0_A0_B\ra, |1_A1_B\ra, |1_A0_B\ra, |0_A1_B\ra)$, or the non-local 
maximally entangled Bell-states\cite{aa,aav,BGNP}.
Other operators with non-maximally entangled eigenstates
like
\beq
\alpha |0_A0_B\ra + \beta |1_A1_B\ra
\eeq
$|\alpha|\ne |\beta|$, and even the ``local'' operators
with non-entangled one-side ``twisted'' eigenstates basis
\beq
|0_A\ra |1_B\ra, \ \
|0_A\ra 0_B\ra, \ \
|1_A\ra(|0_B\ra+ |1_B\ra), \ \
|1_A\ra(|0_B\ra- |1_B\ra)
\eeq
cannot be measured without violating causality.
The situation becomes even more disturbing for higher dimensional
bipartite systems. For a $4\times4$ it has been argued that
certain {\em causal} operators are in fact not measurable\cite{BGNP}.

The purpose of this article is to show that the limitations
described above can be avoided by relaxing the requirements that
measurements needs to satisfy.
The key point is that ideal quantum measurements
play a dual role. They serve us to {\em observe} an unknown quantity, and
to {\em prepare} an initial state of the system.
Indeed the above mentioned causal restrictions
on measurability, have been derived under the assumption
that after a measurement is completed,
the system collapses to an eigenstates of the
observable\cite{Pdef}.
  This preparation assumption  makes it simpler to verify the result
of a measurement by repeating again the measurement.
However the roles of observation and preparation,  are
logically independent.
In the present article we hence assume that:

\vspace {0.5 cm}
\noindent
{\em
A measurement of an observable
does not necessarily prepare eigenstates of
that same observable.
}
\vspace {0.5 cm}

Instead our measurement procedure will prepare either one of the 
Bell-states or, with equal unbiased probabilities, one 
of the untwisted direct product states.
Hence, in accordance with causality\cite{pv}, 
in both cases the local information is erased, and 
the local reduced mixed state is a unit matrix.  
This turns out not only to removes the conflict with causality,
but also allows us to construct measurements for non-trivial cases.
In fact, we show that at least for $2\times2$ dimensional bipartite
systems {\em all} the observable can be measured.

Other requirements from the measurement remain intact:
1) When the initial state is an eigenstate of the measured observable,
the outcome is with certainty given by the corresponding eigenvalue.
2) For a general state, the corresponding eigenvalues are observed with 
the usual quantum probability.
3) The measurement is  instantaneous in the sense that
the local interactions with measuring devices is completed
in time $\Delta t$, that can be taken to be arbitrarily small.
4) The measurement is performed locally at two space-time regions that
are causally disconnected.  Particularly $c\Delta t\ll L$, where  $ L$ is the
distance between the systems A and B.
5) The outcome of the measurement
is recorded locally by the observers at A and B,
and becomes known only after the local recordings are compared.
Finally, we  assume that the resources used
to perform the measurement include an unlimited supply of
entangled pairs which have been earlier distributed between
the parties. We will see that different operators require different
entanglement resources.

In order to implement a measurement we use the following method.
We perform a remote instantaneous transformation of
the unmeasurable set of eigenstates to a measurable
set. The map between the sets is not deterministic.
Different maps can be generated.  However in all cases, we can
use the local outcomes of the measurements, to
infer the relation between the unmeasurable
states and the new measurable states. On the latter states we can
then perform ordinary measurements.
The final result is then obtained by combining the information
on the map, with the local measurement outcomes.

We will demonstrate our method on a number of examples.
In section 2. and 3. we consider $2\times 2$ twisted product 
basis.
In section 4. we apply our method to the case of non-maximal
eigenstates, and in section 5. to the  twisted set\cite{BGNP} in a  $4\times 4$
dimensional Hilbert space.


\section{$2\times2$ twisted product basis}\label{os}

Consider a $2 \times 2$ Hilbert space of a bipartite system,
held by Alice and Bob, and spanned by the local basis vectors
$|0_A\ra$, $|1_A\ra$ and
$|0_B\ra$, $|1_B\ra$, respectively.
In the forthcoming, we shall handle the basis $|0\ra$, $|1\ra$
as spin $\frac{1}{2}$-like states, $|0\ra\equiv|\uparrow_z\ra$,
$|1\ra\equiv|\downarrow_z\ra$, and define accordingly the Pauli operators.

The first example which we discuss is a non-degenerate operator,
defined by the four mutually orthogonal direct-product eigenstates:
\begin{eqnarray}\label{4st}
|\Psi_{AB}^1\rangle=&|0_A\rangle&|0_B\rangle,\nonumber\\
|\Psi_{AB}^2\rangle=&|0_A\rangle&|1_B\rangle,\\
|\Psi_{AB}^3\rangle=&|1_A\rangle~&{\scriptstyle\frac{1}{\sqrt2}}(|0_B\rangle+|1_B\rangle),\nonumber\\
|\Psi_{AB}^4\rangle=&|1_A\rangle~&{\scriptstyle\frac{1}{\sqrt2}}(|0_B\rangle-|1_B\rangle),\nonumber
\end{eqnarray}
\noindent
This operator was shown to be semilocalizable \cite{BGNP,gv},
i.e. it can be measured with the assistance of one-directional classical
communication from Alice to Bob.  We now show that
this operator may be measured {\em instantaneously}, without any
exchange of information, by utilizing one ebit of shared entanglement.

The main idea here, is to perform a conditional
$\pi/2$ rotation of Bob's state when Alice's state is $|1_A\ra$.
This maps the twisted basis on Bob's side
\beq
 \lbrace |0_B\rangle+|1_B\rangle, \ |0_B\rangle-|1_B\rangle \rbrace
\ \ \longrightarrow \ \  \lbrace |0_B\ra, |1_A\ra \rbrace
\eeq
On the other hand, no such mapping will be performed if Alice's
state turns out to be $|0_A\ra$.

The "switch" which controls the remote rotation
is in Alice's hands.
Bob on the other hand will performs a fixed set of
manipulations independently of Alice's choice, and finally measures
$\sigma_{z_B}$. Having done so, Alice and
Bob obtain  local records that when combined allow us to
distinguish with certainty between the elements of the basis
(\ref{4st}).
The final state of Alice's and Bob's spin is always given by a unit density
matrix. Hence, although the whole process takes place instantaneously,
no violations of causality occur.

Let now consider the process in some detail.
Several methods for generating remote operations 
have been suggested\cite{popescu,plenio,huelga,stator,levcomment}.
Here we follow the state-operator (stator) method of Ref.  \cite{stator},
which allows simple and intuitive construction of a large class of
remote operations.
For the present case, Alice and Bob need one shared ebit
in order to perform a remote rotation. The initial state at the
hands of Alice and Bob is then
\beq
{1\over\sqrt2}\biggl(|0_a\ra\otimes|0_b\ra + |1_a\ra\otimes
|1_b\ra \bigg)\otimes  |\Psi_{AB}\ra\ \eeq
\noindent

\noindent where  by the small letters, $a$ and $b$, we denote the
ancillary ebit, shared by Alice and Bob, respectively.
Bob starts by performing a local CNOT interaction (with respect to
$\sigma_{y_B}$) between the entangled qubit $b$ and his state $B$,
described by the unitary transformation
\beq\label{cnot} U_{bB} =
|0_b\ra\la 0_b| \otimes I_{B} + |1_b\ra\la1_b|\otimes \sigma_{y_B}
\eeq

\noindent This yields the state

\beq {1\over\sqrt2}\biggl(|0_a\ra\otimes|0_b\ra\otimes I_{B}
 + |1_a\ra\otimes |1_b\ra \otimes\sigma_{y_B} \biggr)|\Psi_{AB}\ra
\eeq

Next he performs a measurement of $\sigma_{x_b}$ of the
 entangled qubit $b$ and keeps the result $v(\sigma_{x_b})$.
The resulting state is now
\beqa\label{resx}
 \biggl(|0\ra_a\otimes I_B \pm |1\ra_a\otimes\sigma_{y_B} \biggr)
|\Psi_{AB}\ra \nonumber \\
\equiv {\cal S} |\Psi_{AB}\ra
\eeqa

\noindent where the $\pm$ above corresponds to $v(\sigma_{x_b})$.
The object ${\cal S}$ is the state-operator (stator)
defined in \cite{stator}. The stator satisfies
the eigenoperator equation:
\beq
\sigma_{x_a} {\cal S} = v(\sigma_{x_b})\sigma_{y_B}{\cal S}
\eeq
This equation captures the correlations between a unitary transformation
which is acted by Alice on $a$ and the equivalent rotation
on an arbitrary state of Bob.
Particularly, we also have that if Alice acts with the
unitary transformation $\exp (i\alpha \sigma_{x_a})$, that is
equivalent to the unitary transformation $\exp (i\alpha \sigma_{y_B})$
on Bob's qubit (up to a trivial $\pi$ rotation around the $y$-axis).

Having "prepared" the above stator,
 Alice proceeds and measures $\sigma_{z_A}$.
 Consider the two possible outcomes. If it turns out
 that $|\Psi_A\ra= |0_A\ra$, she next measures  $\sigma_{z_a}$
 and keeps the result $v( \sigma_{z_a})$.
This will induce on Bob's qubit the transformation
\beq
{1+v( \sigma_{z_a}) \over 2}1_B + v(\sigma_{x_b}){1-v( \sigma_{z_a})\over 2}
\sigma_{y_B}
\eeq
If, on the other hand, she gets $|\Psi_A\ra= |1_A\ra$,
she acts by $e^{i{\pi\over4}
\sigma_{x_a}}$ on $a$ in order to untwist Bob's qubit,
and then measures $\sigma_{z_a}$ of  $a$.
The induced remote transformation is in this case
\beq
\biggl( {1+v( \sigma_{z_a}) \over 2}1_B +v(\sigma_{x_b})
 {1-v( \sigma_{z_a})\over 2}
\sigma_{y_B} \biggr) e^{{i\pi\over 4} v(\sigma_{x_b}) \sigma_{y_B}}
\eeq
The four possible outcomes of this map are summarized in the
following table:

\begin{center}

\begin{tabular}{|c|c|c|}   \hline

$\sigma_{z_a}\backslash\sigma_{x_b}=+1$&$v(\sigma_{x_b})=+1$&$v(\sigma_{x_b})=-1$\\ \hline

$v(\sigma_{z_a})=+1$&$|\Psi_{AB}^1\rangle \rightarrow |0_A\ra|0_B\ra$&$|\Psi_{AB}^1\rangle\rightarrow|0_A\ra|0_B\ra$\\
&$|\Psi_{AB}^2\rangle\rightarrow|0_A\ra|1_B\ra$&$|\Psi_{AB}^2\rangle\rightarrow|0_A\ra|1_B\ra$\\
&$|\Psi_{AB}^3\rangle\rightarrow|1_A\ra|0_B\ra$&$|\Psi_{AB}^3\rangle\rightarrow|1_A\ra|1_B\ra$\\
&$|\Psi_{AB}^3\rangle\rightarrow|1_A\ra|1_B\ra$&$|\Psi_{AB}^3\rangle\rightarrow|1_A\ra|0_B\ra$\\
\hline

$v(\sigma_{z_a})=-1$&$|\Psi_{AB}^1\rangle\rightarrow|0_A\ra|1_B\ra$&$|\Psi_{AB}^1\rangle\rightarrow|0_A\ra|1_B\ra$\\
&$|\Psi_{AB}^2\rangle\rightarrow|0_A\ra|0_B\ra$&$|\Psi_{AB}^2\rangle\rightarrow|0_A\ra|0_B\ra$\\
&$|\Psi_{AB}^3\rangle\rightarrow|1_A\ra|1_B\ra$&$|\Psi_{AB}^3\rangle\rightarrow|1_A\ra|0_B\ra$\\
&$|\Psi_{AB}^4\rangle\rightarrow|1_A\ra|0_B\ra$&$|\Psi_{AB}^4\rangle\rightarrow|1_A\ra|1_B\ra$\\
\hline

\hline

\end{tabular}

\end{center}

To complete the measurement of the semi-local set, Bob measures
the operator $\sigma_{z_B}$ for his state $|\Psi_B\ra$.
Finally, in order to infer which eigenstate did they measure,
Alice and Bob use $v(\sigma_{z_a})$  and
$v(\sigma_{x_b})$ to identify which of the above four blocks (maps)
has been realized, and then use the results for $\sigma_{z_A}$ and
$\sigma_{z_B}$ to isolate one of the four possible states in that block.

\section{General $2\times2$ product basis}\label{4gset}
 We can implement our method to the more general case by applying
 a general rotation to Bob's qubit spin in (\ref{4st}):
\begin{eqnarray}
\label{4g}
|\Psi^1_{AB}\rangle=&|0_A\rangle&|0_B\rangle,\nonumber\\
|\Psi^2_{AB}\rangle=&|0_A\rangle&|1_B\rangle\\
|\Psi^3_{AB}\rangle=&|1_A\rangle&(\cos{\alpha\over2}|0_B\rangle+\sin{\alpha\over2}|1_B\rangle)\nonumber,\\
|\Psi^4_{AB}\rangle=&|1_A\rangle&(\sin{\alpha\over2}|0_B\rangle-\cos{\alpha\over2}|1_B\rangle)\nonumber,
\end{eqnarray}
with $0 <\alpha < {\pi}$. For simplicity, and without loss of generality,
we have ignored possible relative phases in $|\Psi^3_{AB}\ra$ and
$|\Psi^4_{AB}\ra$. 

Now our task is more complicated, because the
rotation of $|\Psi_B^3\rangle$ and
$|\Psi_B^4\rangle$ performed in the previous section succeeds here
only with probability 1/2 to map Bob's  set to  $|0_B\rangle$ and
$|1_B\rangle$. With probability 1/2 the transformation takes us
to $(\cos{\alpha}|0_B\rangle+\sin{\alpha}|1_B\rangle)$ and
$(\sin{\alpha}|0_B\rangle-\cos{\alpha}|1_B\rangle)$.
Hence we fail for the case
$\alpha\neq\frac{\pi}{2}$.
Fortunately, the information about which map has been realized
is available to Bob. 
If $v(\sigma_{x_b})=+1$, Bob knows
that they succeeded. If the result was $v(\sigma_{x_b})=-1$,
Alice and Bob can perform an additional rotation with an angle
$2\alpha$ in order correct the error.
This second step again succeeds  with
probability 1/2. Thus the total probability of success is now 3/4, 
and the process can be repeated again until the probability of failure 
is sufficiently small.
This method of corrections has been proposed in Ref.\cite{vidal+cirac}.
The correction steps require the supply of additional
ebits. For the general case, $n$ ebits allow us to perform a successful
measurement with probability $1-\frac{1}{2^{n-1}}$. 
The ebits are  necessary even if the process truncates
after a finite small number of steps once Bob obtained $v(\sigma_{x_b})=+1$.
Since we do not allow any communication,
Alice must apply a fixed set of rotations 
$e^{i(n-1)\alpha\sigma_{z_a}}$ on the $n$'th ebit.

Nevertheless, not for all angles the infinite number of ebits
is needed in order to get probability infinitesimally  close  to one.
If $\alpha=\frac{\pi k}{2^{n+1}}$ then in the n+1 step they will always
succeed. Thus, for certain angles the number of ebits $n$ is
relatively small: $n=2$ for $\alpha=\frac{\pi}{8},\frac{3\pi}{8}$,
$n=3$ for $\alpha=\frac{\pi}{16},\frac{3\pi}{16},\frac{5\pi}{16}$,
etc.

\section{Nonmaximal entangled eigenstates}
The operator with four non-degenerate Bell-state eigenstates is known
to be measurable instantaneously \cite{aa,aav,BGNP}.
However, for the case of ideal measurements, an observable  
 with the nonmaximal eigenstates,
\begin{eqnarray}\label{nmax}
|\Psi_{AB}^1\rangle=\cos{\alpha\over2}|0_A0_B\ra+\sin{\alpha\over2}
|1_A1_B\ra,\nonumber\\
|\Psi_{AB}^2\rangle=\sin{\alpha\over2}|0_A0_B\ra-\cos{\alpha\over2}|1_A1_B\ra,\\
|\Psi_{AB}^3\rangle=\cos{\alpha\over2}|0_A1_B\ra+\sin{\alpha\over2}
|1_A0_B\ra,\nonumber\\
|\Psi_{AB}^4\rangle=\sin{\alpha\over2}|0_A1_B\ra-\cos{\alpha\over2}
|1_A0_B\ra,\nonumber
\end{eqnarray}
it was shown to be inconsistent with causality. 
Here we show that our method allows a precise measurement of
a non-maximal operator with the above eigenstates. We later
consider the somewhat more complicated case of non-maximal
eigenstates with non-equal entanglement.

The measurement has two main steps.
First, one ebit is used to perform a remote CNOT, which acts to
transform the above set, to a set of mutually orthogonal
non-entangled product states, reminiscent to the twisted set
considered in section 3.
Then, in the second step,  the twisted set, is ``untwisted''
by a remote rotation. After this step Alice and Bob can
apply  ordinary measurements on their qubits.

Let us spell-out the details of the process.
Alice and Bob begin as before by preparing the stator (\ref{resx}).
Then Alice generates remote CNOT-like transformation
on Bob's qubit by applying the local unitary transformation
$\exp(-i{\pi\over4}(1-\sigma_{z_A})(1 - \sigma_{x_a} ))$
on the half $a$ of the ebit and her qubit $A$, followed by a
measurement of $\sigma_{z_a}$.
This activates on Bob's qubit the unitary transformation
\beq\label{cnot1}
\biggl( {1+ v(\sigma_{z_a}) \over 2} 1_B + v(\sigma_{x_b}){1-v(\sigma_{z_a})\over 2}
\sigma_{x_B}\biggr) e^{-i\frac{\pi}{4}(1-\sigma_{z_A})(1 - v(\sigma_{x_b}) \sigma_{x_B} ) }.\\
\label{UAB}
\eeq
Notice that for $v(\sigma_{z_a})=v(\sigma_{x_b})=1$ that is the
usual CNOT transformation. Otherwise we have to incorporate some
trivial corrections. In the case $v(\sigma_{z_a})=+1$, the state transforms to

\begin{eqnarray}
|\Psi_{AB}^1\rangle=(\cos{\alpha\over2}|0_A\ra +
v(\sigma_{x_b})\sin{\alpha\over2}|1_A\ra)|0_B\ra,\nonumber\\
|\Psi_{AB}^2\rangle=(\sin{\alpha\over2}|0_A\ra - v(\sigma_{x_b})
\cos{\alpha\over2}|1_A\ra)|0_B\ra,\\
|\Psi_{AB}^3\rangle=(\cos{\alpha\over2}|0_A\ra + 
v(\sigma_{x_b})\sin{\alpha\over2}|1_A\ra)|1_B\ra,\nonumber\\
|\Psi_{AB}^4\rangle=(\sin{\alpha\over2}|0_A\ra - 
v(\sigma_{x_b}) \cos{\alpha\over2}|1_A\ra)|1_B\ra,\nonumber
\end{eqnarray}
 If Alice gets $\sigma_{z_a}=-1$, the set is similar with Bob's
spin is flipped.

At this point, Alice or Bob still cannot extract any information
because the directions are still ``twisted''.
However Alice's qubit is in either of the two orthogonal
states of a spin that has been rotated by an angle $v(\sigma_{x_b})
\alpha$ around the y-axis.
To untwist the set,  Bob now performs a remote rotation of Alice's
spin by the angle
 $v(\sigma_{x_b})\alpha$.
As before, the number of ebits needed to utilized this step depends on
 the angle $\alpha$, but with  unlimited ebit resources we can always
 succeed with  probability arbitrary close to unity.
After completing this step (assuming that we succeeded)
we obtain the following four maps  summarized in the table:

\begin{center}

\begin{tabular}{|c|c|c|}   \hline

$\sigma_{z_a} \backslash \sigma_{x_b}$&$v(\sigma_{x_b})=+1$&$v(\sigma_{x_b})=-1$\\              \hline

$v(\sigma_{z_a})=+1$&$|\Psi_{AB}^1\rangle\rightarrow|0_A\ra|0_B\ra$&$|\Psi_{AB}^1\rangle\rightarrow|0_A\ra|0_B\ra$\\
&$|\Psi_{AB}^2\rangle\rightarrow|1_A\ra|0_B\ra$&$|\Psi_{AB}^2\rangle\rightarrow|1_A\ra|0_B\ra$\\
&$|\Psi_{AB}^3\rangle\rightarrow|0_A\ra|1_B\ra$&$|\Psi_{AB}^3\rangle\rightarrow|0_A\ra|1_B\ra$\\
&$|\Psi_{AB}^4\rangle\rightarrow|1_A\ra|1_B\ra$&$|\Psi_{AB}^4\rangle\rightarrow|1_A\ra|1_B\ra$\\
\hline
$v(\sigma_{z_a})=-1$&$|\Psi_{AB}^1\rangle\rightarrow|0_A\ra|1_B\ra$&$|\Psi_{AB}^1\rangle\rightarrow|0_A\ra|1_B\ra$\\
&$|\Psi_{AB}^2\rangle\rightarrow|1_A\ra|1_B\ra$&$|\Psi_{AB}^2\rangle\rightarrow|1_A\ra|1_B\ra$\\
&$|\Psi_{AB}^3\rangle\rightarrow|0_A\ra|0_B\ra$&$|\Psi_{AB}^3\rangle\rightarrow|0_A\ra|0_B\ra$\\
&$|\Psi_{AB}^4\rangle\rightarrow|1_A\ra|0_B\ra$&$|\Psi_{AB}^4\rangle\rightarrow|1_A\ra|0_B\ra$\\
\hline

\end{tabular}

\end{center}

The information about columns
is located only on the Bob's side, while information about rows is
located on Alice's side. Finally Alice and Bob measure $\sigma_z$ of
their qubit $A$ and $B$ and by comparing results can identify
the relevant non-maximal eigenstate.

The final state of the qubits $A$ and $B$ is in this case one of the equally
probable direct product states. It is interesting to note that one 
can also perform, 
a measurement which achives the same goal, but collapses the 
state to one of the Bell-states. 
To this end, we notice that the basis (\ref{nmax}) can also be written
as 
\begin{eqnarray}
|\Psi_{AB}^1\rangle=U_A |0_A(0+1)_B\ra+U_A^\dagger|1_A(0-1)_B\ra,\nonumber\\
|\Psi_{AB}^2\rangle=U_A|1_A(0+1)_B\ra+U^\dagger_A|0_A(0-1)_B\ra,\\
|\Psi_{AB}^3\rangle=U_A|1_A(0+1)_B\ra-U^\dagger_A|1_A(0-1)_B\ra,\nonumber\\
|\Psi_{AB}^4\rangle=U_A|0_A(0+1)_B\ra-U^\dagger_A|0_A(0-1)_B\ra,\nonumber
\end{eqnarray}
where $U_A=\exp({i{\alpha\over 2}\sigma_{y_A}})$. Hence Bob can perform 
a controlled rotation in order to undo the unitary transfomation 
$U_A$ and $U_A^\dagger$. To complete the measurement, Alice and Bob 
need two more ebits in order to distinguish between the four
Bell states. 
The resources needed for the this
method require one more ebit.  However, the final state of the system 
is in this case a maximall entangled state, therefore both methods
actually consume
the same amount of entanglement.
As in the previous case, the reduced density operator of a single qubit
is given by a unit matrix.

Our scheme works also for the most general operators with
entangled eigenstates.
\begin{eqnarray}\label{nmax2}
|\Psi_{AB}^1\rangle=\cos{\alpha\over 2}|0_A0_B\ra+e^{i\phi_1}\sin
{\alpha\over 2}|1_A1_B\ra,\nonumber\\
|\Psi_{AB}^2\rangle=\sin{\alpha\over2}|0_A0_B\ra-
e^{i\phi_1}\cos{\alpha\over2}|1_A1_B\ra,\\
|\Psi_{AB}^3\rangle=\cos{\beta\over2}|0_A1_B\ra+e^{i\phi_2}\sin
{\beta\over2}|1_A0_B\ra,\nonumber\\
|\Psi_{AB}^4\rangle=
\sin{\beta\over2}|0_A1_B\ra-e^{i\phi_2}\cos{\beta\over2}|1_A0_B\ra,\nonumber
\end{eqnarray}
In the following we can set the  phases $\phi_1=\phi_2=0$ without
loss of generality. 
\noindent The initial step remains the same.
Alice and Bob apply a remote
CNOT. However in the next step,  Bob has to apply one of  two 
possible remote  rotations on Alice's qubit. 
If his local qubit is $|0_B\ra$ he rotates Alice's qubit, with respect 
to the $y$ axis, by
$v(\sigma_{x_b})\alpha$. When his local qubit is $|1_B\ra$ he rotates
Alice's qubit by $v(\sigma_{x_b})\beta$. 
(More generally, for non-vanishing $\phi_1$ and $\phi_2$ 
the axes of rotation must be chosen appropriately.) 
After this step the four possible
sets become:

\begin{center}
\begin{tabular}{|c|c|c|}   \hline

$\sigma_{z_a} \backslash
\sigma_{x_b}$&$v(\sigma_{x_b})=+1$&$v(\sigma_{x_b})=-1$\\ \hline

$v(\sigma_{z_a})=+1$&$~|0_A\ra|0_B\ra$&$~|0_A\ra|0_B\ra$\\
&$~|1_A\ra|0_B\ra$&$~|1_A\ra|0_B\ra$\\
&$~|0_A\ra|1_B\ra$&$~|0_A\ra|1_B\ra$\\
&$~|1_A\ra|1_B\ra$&$~|1_A\ra|1_B\ra$\\
\hline
$v(\sigma_{z_a})=-1$&$(\cos \gamma |0_A\ra+\sin \gamma |1_A\ra)|1_B\ra$&$(\cos \gamma |0_A\ra-\sin \gamma |1_A\ra)|1_B\ra$\\
&$(\sin \gamma |0_A\ra-\cos \gamma |1_A\ra)|1_B\ra$&$(\sin \gamma |0_A\ra+\cos
\gamma |1_A\ra)|1_B\ra$\\
&$(\cos \gamma |0_A\ra-\sin \gamma |1_A\ra)|0_B\ra$&$(\cos \gamma
|0_A\ra+\sin \gamma |1_A\ra)|0_B\ra$\\
&$(\sin \gamma |0_A\ra+\cos \gamma |1_A\ra)|0_B\ra$&$(\sin \gamma
|0_A\ra-\cos \gamma |1_A\ra)|0_B\ra$\\
\hline

\end{tabular}
\end{center}
where $\gamma=(\alpha-\beta)/2$.
\noindent The two upper blocks are obtained for $v(\sigma_{a_z})=+1$
while the lower row for $v(\sigma_{a_z})=-1$.
For the latter case, an additional rotation on the angle
$\beta-\alpha$  is needed 
in order to rotate the four sets from second row to the $z$-direction.
To obtain this, Alice and Bob use additional
entanglement. Alice engages Bob's particle into this transformation 
only if $v(\sigma_{z_a})=-1$.
Bob determines the rotation angle according to his previous outcomes
as follows:
if $v(\sigma_{x_b})v(\sigma_{x_B})=1$ he rotates Alice's qubit by
$\beta-\alpha$.  If $v(\sigma_{x_b})v(\sigma_{x_B})=-1$
he rotates by $\alpha-\beta$.
Thus, finally we arrive to the sets:

\begin{center}
\begin{tabular}{|c|c|c|}   \hline

$\sigma_{z_a} \backslash
\sigma_{x_b}$&$v(\sigma_{x_b})=+1$&$v(\sigma_{x_b})=-1$\\ \hline

$v(\sigma_{z_a})=+1$&$|0_A\ra|0_B\ra$&$|0_A\ra|0_B\ra$\\
&$|1_A\ra|0_B\ra$&$|1_A\ra|0_B\ra$\\
&$|0_A\ra|1_B\ra$&$|0_A\ra|1_B\ra$\\
&$|1_A\ra|1_B\ra$&$|1_A\ra|1_B\ra$\\
\hline
$v(\sigma_{z_a})=-1$&$|0_A\ra|1_B\ra$&$|0_A\ra|1_B\ra$\\
&$|1_A\ra|1_B\ra$&$|1_A\ra|1_B\ra$\\
&$|0_A\ra|0_B\ra$&$|0_A\ra|0_B\ra$\\
&$|1_A\ra|0_B\ra$&$|1_A\ra|0_B\ra$\\
\hline

\end{tabular}
\end{center}

\noindent
The full process is summarized in Figure 1. 

\begin{figure} \epsfysize=4truein
      \centerline{\epsffile{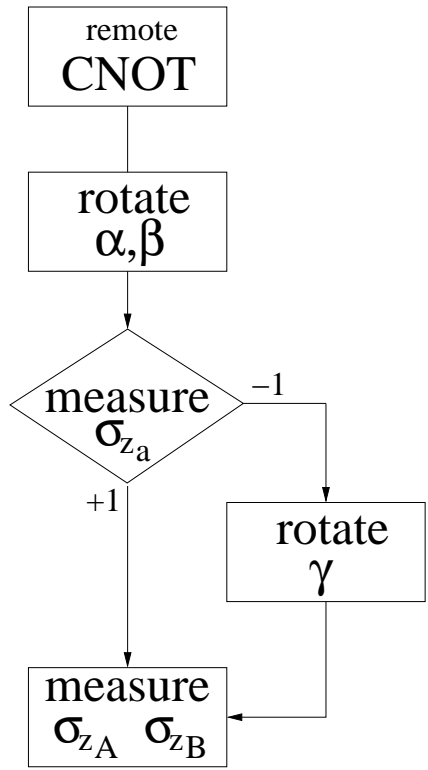}}
\vspace {0.5cm}
  \caption[]{Schematic plot of the steps in 
general case of non-maximal entangled eigenstates}
    \label{fig1} \end{figure}

\eject
\section{$4\times4$ Twist}
Another interesting example of a measurable nonlocal operator is the  twisted
operator\cite{BGNP} living in $4\otimes4$ dimensional Hilbert
space. This operator is defined by the following 16 entangled
eigenstates:

\begin{center}
\begin{tabular}{|c|c|c|}   \hline

&$0_B1_B$&$2_B3_B$\\         \hline
$0_A1_A$
&$|0_A0_B\ra\pm|1_A1_B\ra$&$|0_A2_B\ra\pm|1_A3_B\ra$\\
&$|0_A1_B\ra\pm|1_A0_B\ra$&$|0_A3_B\ra\pm|1_A2_B\ra$\\
 \hline

$2_A3_A$
&$|2_A0_B\ra\pm|3_A1_B\ra$&$U_B(|2_A2_B\ra\pm|3_A3_B\ra)$\\
&$|2_A1_B\ra\pm|3_A0_B\ra$&$U_B(|2_A3_B\ra\pm|3_A2_B\ra)$\\
 \hline
\end{tabular}
\end{center}

\noindent where  $U_B$ is some unitary
rotation operator acting on the state of Bob's particle. Alice's
and Bob's Hilbert spaces are  partitioned to two
two-dimensional subspaces. Each quadrant contains four Bell
states. If we can perform the transformation $U_B^{\dagger}$ on the
$2_A3_A\times 2_B3_B$-quadrant, then the original set transforms
to a measurable basis. To this end, Alice
and Bob perform a measurement to project the local states of their
particles into the subspaces $0_A$,$1_A$ or $2_A,3_A$ and $0_B$,$1_B$
or $2_B,3_B$. As a result the initial 16-set collapses to a 4-state
set. Of course, the information about it is distributed between
the parties: Alice knows the row, Bob knows the column.
But we have encountered a very similar situation in the case of 
discussed in section 3. 
Hence Alice and Bob can untwist this transformation 
using one shared ebit. 
Bob performs a CNOT interaction (\ref{cnot}) only if his state was
projected to $2_B,3_B$ subspace, and Alice acts on her qubit $a$
with an appropriate transformation $U_a$ (which generates 
acting on the stator $U_B^\dagger$)
only if her state was projected into the subspace $2_A,3_A$. 
To complete the measurement Alice and Bob perform a remote measurement
of the Bell operator.


\section{Summary}

We have reconsidered the
measurability problem for nonlocal operators which have been previously  
considered to be in conflict with relativistic causality and hence
unmeasurable. 
We argued that, if the preparation role
of an ideal measurement is relaxed, some observables do become
measurable.
In fact we showed  that {\em every} bipartite operator
in $2\otimes 2$ Hilbert space can be 
measured instantaneously. We also demonstrated  that the  
operator discussed in \cite{BGNP} for a bipartite $4\otimes 4$ 
dimensional system is measurable in this sense.
The approach suggested here, seems to  broaden considerably 
the class of measurable observables. 
However, the general question of measurability is still open.
It remains to be seen, whether similar methods are applicable 
for general bipartite and multi-particle cases.
 
\vspace{.2cm}

We thank Y. Aharonov and L. Vaidman for useful discussions.
The research was supported in part by grant 62/01-1 of
the Israel Science Foundation, established by the
Israel Academy of Sciences and Humanities.


\begin{thebibliography}{9}
\bibitem{Pdef} In Ref. \cite{BGNP} this definition is replaced by a weaker one:
the measurement decoheres the density matrix according to the measured
quantity, and no local record of the measurement outcome is necessarily
produced. Nevertheless the resulting restriction on the set of observables 
remains intact.
\bibitem{levcomment} L. Vaidman
has shown that the instantaneous remote rotations needed for the 
present article may also be produced by sequence of teleportation steps.
\end{thebibliography}
\end{document}